\documentclass[preprint,pra,aps,prfluids,showkeys,superscriptaddress,longbibliography]{revtex4-2}

\usepackage{color}
\usepackage[caption=false]{subfig}

\usepackage{amsmath,amssymb,amsfonts} %
\usepackage{graphicx}

\usepackage{multirow}

\usepackage{color}

\begin{document}


\title{Is it possible to overheat ice? The activated melting of TIP4P/Ice at solid-vapor coexistence.}

\author{
	Lukasz Baran\textsuperscript{a}, 
	Pablo Llombart\textsuperscript{b},
    Eva G. Noya\textsuperscript{c} and 
    L. G.~MacDowell\textsuperscript{d}
    \email[]{lgmac@quim.ucm.es}
\affiliation{
\textsuperscript{a} Department of Theoretical Chemistry, Institute of Chemical Sciences, Faculty of Chemistry, Maria Curie-Sklodowska University in Lublin, Lublin, Poland.\\
\textsuperscript{b} Department of Theoretical Condensed Matter Physics, Condensed Matter Physics Center (IFIMAC) and Instituto Nicol{\'a}s Cabrera, Universidad Aut{\'o}noma de Madrid, 28049, Spain.\\
\textsuperscript{c} Instituto de Qu\'{\i}mica F\'{\i}sica Blas Cabrera, CSIC, Calle Serrano 119, 28006 Madrid, Spain.\\
\textsuperscript{d} Departamento de Qu\'{\i}mica-F\'{\i}sica (Unidad Asociada de I+D+i al CSIC),
Facultad de Ciencias Qu\'{\i}micas, Universidad Complutense de Madrid, 28040 Madrid, Spain.
}
}

\begin{abstract}
	A widely accepted phenomenological rule states that solids with free surfaces cannot be overheated. In this work we discuss this statement critically under the light of the statistical thermodynamics of interfacial roughening transitions.   Our results show that the basal face of ice as described by the TIP4P/Ice model can remain mechanically stable for more than one hundred nanoseconds when overheated by 1~K, and for several hundreds of nanoseconds at smaller overheating despite the presence of a significant quasi-liquid layer at the surface. Such time scales, which are often of little experimental significance, can become a concern for the determination of melting points by computer simulations using the direct coexistence method. In the light of this observation, we reinterpret computer simulations of ice premelting and show that current results for the TIP4P/Ice model all imply a scenario of incomplete surface melting. Using a thermodynamic integration path, we reassess our own estimates for the Laplace pressure difference between water and vapor. These calculations are used to measure the  disjoining pressure of premelting liquid films and allow us to confirm a minimum of the interfacial free energy at finite premelting thickness of about one nanometer. 
\end{abstract}

\keywords{
   Surface Melting; Premelting; Quasi-liquid layer; Roughening; Layering;  
}

\maketitle

\section{Introduction}

One of the most salient features of first order phase transitions is the presence of a large free energy barrier separating
the coexisting bulk phases \cite{binder87}. Thus, it is not unusual to find a liquid phase that can be overheated or a vapor phase that can be under-cooled beyond the vaporization temperature \cite{debenedetti96}. 

Overheating of a solid phase beyond its melting point, on the contrary, is hardly ever observed. This was acknowledged early in the 20th century by Tamman, Bridgman, Stransky and Frenkel (c.f. \cite{dash99} for a historical account), and has become widely accepted as  an almost universal feature of solids. Of course, the melting transition is first order, but it is thought that at the solid interface, melting occurs instantaneously without any activation as soon as the solid is slightly overheated \cite{nenow84,frenken85,nozieres92,dash95,dash06,mei07}.

A quantification of this observation is provided by the Wilson-Frenkel growth law that is widely used in computer simulation studies to model melting. This assumes a symmetrical mechanism of freezing/melting rates that is linear in the deviations away from the melting temperature, $T_m$. Whence, as soon as $T>T_m$, a crystal starts melting at its surface, and no overheating can be observed in practice \cite{frenken85}.

The situation is actually somewhat more complex than implied in the linear growth mechanism postulated in the Wilson-Frenkel law. In practice, solid/fluid interfaces come in two widely different flavors, rough and smooth, depending on the substance and temperature under study \cite{bienfait92}.  At the microscopic level, the distinction between the low temperature smooth surface and the high temperature rough interfaces is a  statement about the nature of the surface height fluctuations \cite{chaikin95,nelson04,balibar05}. Smooth interfaces exhibit only very small surface height fluctuations, which remain bound to the underlying crystal lattice. Rough interfaces, on the contrary, exhibit large surface height fluctuations that become unbound to the crystal lattice and have similar properties as fluid/fluid interfaces. 

This apparently sophisticated distinction has extremely important  macroscopic manifestations. 

Rough interfaces have a barrier-less growth law that is linear in the driving force, as in the Wilson-Frenkel equation \cite{chui78,weeks79,saito80,bienfait92}. On the contrary,  smooth surfaces in perfect crystals grow only in a  layer by layer fashion via a slow and activated two dimensional growth mechanism  proportional to $\exp(-T_m/(T-T_m))$, whence, at vanishingly small  rate for small but finite overheating \cite{chui78,weeks79,saito80}. 

Furthermore,  macroscopic crystals  surfaces that are smooth appear as flat facets which terminate at sharp edges, while rough interfaces  are rounded and do not exhibit edges \cite{rottman84,bienfait92,miracle-sole13}. Thus, the most straightforward method to display the distinction between smooth and rough facets is the slow heating of an equilibrium crystal: as the temperature increases, facets can eventually become rounded at the {\em roughening temperature}, $T_R$ \cite{balibar05}. But the roughening transition is highly anisotropic and very much depends on the crystal plane \cite{pluis87,tartaglino05}. Low index planes are the most reluctant to roughening, and often remain smooth at all temperatures along the coexistence line \cite{rottman84,pluis87,metois89,nozieres92}.

The important distinction between rough and smooth surfaces, while anticipated by Burton, Cabrera and Frank in the 1950s \cite{burton51} was only worked out in full detail during the 1980s \cite{vanbeijeren77,chui78,weeks79,saito80,rottman84,bienfait92}, and has yet not sufficiently well pervaded the simulation community.

However, it perspires from the above discussion that the barrier-less melting of a crystal can only occur at {\em rough} interfaces. A perfect crystal bound only by smooth interfaces will be able to withstand overheating for a period equivalent to the time scale of activated growth. In some exceptional cases, this can be a very long and verifiable time scale \cite{spiller82,metois89,nozieres92}. But in most experimental situations, either the presence of rough interfaces, dislocations or grain boundaries will make the observation of overheating very rare, and the time of activation, if any, too short for practical observation. 

An important exception concerns the study of melting by computer simulations. Here, small, defect-less crystals are prepared which usually exhibit interfaces along one single crystallographic direction, and are made pseudo-infinite by virtue of the periodic boundary conditions in the remaining parallel directions \cite{fernandez06,conde17,rozas21,bore22,ji24}. Surface initiated melting then occurs at a single interfacial plane. The nature of the exposed interface, whether rough or smooth will dictate the extent of achievable overheating, a situation that has often not been recognized. This is particularly important for the study of phase boundaries by direct coexistence, since a   microscopically small activation time that can hardly be measured in experiment,  can be in fact several times longer than the full simulation time scale, leading to an overheated crystal which remains stable along the full trajectory of the finite time simulation. 

The question regarding the likelihood of overheating in ice is then a question on the nature of the exposed crystal planes. 

The most frequently observed ice crystal planes  are the
basal (0001) and prism planes $(10\overline{1}0)$, although other orientations can be exposed too \cite{nakaya54,libbrecht22}. At the ice/water interface, the basal plane is thought to remain smooth at all temperatures, a fact supported by direct visual observation of elementary steps on vicinal basal planes \cite{murata22}, as well as by indication of layer-wise growth in computer simulations \cite{nada05,rozmanov12,mochizuki23}. On the contrary, the prism plane has been observed to roughen at about $-16$~$^{\circ}$C \cite{maruyama97,maruyama05}. It is expected that all other crystal planes of ice, being of higher indexes, have become rough at temperatures below the roughening temperature of the prism plane. These set of observations then suggest that only ice crystals exposing their basal facet or their vicinal planes could be overheated above the bulk melting temperature in principle. In practice, experiments of ice grown from supercooled water show that the nucleated regime is observed only in a very small range of about 0.1~K about the melting point, implying a very small barrier for nucleation and very limited possibility of overheating \cite{furukawa21}.

At a first thought, the interface of ice with its vapor is a better candidate for the observation of overheating, as one expects a much sharper interface in view of the large density difference between the vapor and the solid phase. In practice, ice, as most other solids, shows a clear propensity for premelting, i.e., the formation of a so called quasi-liquid layer of molten ice at its surface \cite{kuroda82,nenow84,dash06}. Thus, the sharp ice/vapor transition region is replaced by a broad interface, with a premelted liquid film in between the ice and vapor bulk phases. 

This hypothesis has a long history of controversies (c.f. \cite{weyl51,jellinek62,rosenberg05} for an interesting perspective),  that has been now amply confirmed.  Early evidence was gathered from insightful interpretation of rudimentary experiments  \cite{nakaya54b,hosler57,gilpin80}, and exploited to model the complex structure of ice crystals \cite{lacmann72,kuroda82}, despite criticism \cite{knight96}. This has thereafter gradually built into overwhelming experimental evidence from spectroscopic measurements, \cite{furukawa87,kouchi87,maruyama00,dosch95,wei01,bluhm02,sadtchenko02,smit17,sanchez17},  mechanical probes \cite{doppenschmidt00,constantin18}, direct visual observation \cite{sazaki12,murata16,murata19} and computer simulations \cite{vega06,conde08,neshyba09,limmer14,neshyba16,benet16,sanchez17,kling18,qiu18,benet19,llombart19,llombart20b,berrens22,cui22,cui23,yasuda24}. 

The presence of a premelting film effectively provides ice in coexistence with its vapor  with an ice/liquid interface. This could suggest ice melts from the vapor just as easily as it does from the liquid. Indeed, the ice/vapor  interface also has a basal plane that remains smooth all the way up to the triple point, and a  prism plane that roughens on approaching the melting point. Furthermore,  some evidence suggests the step free energies at the ice/vapor interface could be similar to those found for growth from the melt \cite{mochizuki23,libbrecht14,libbrecht22}, an assumption implicit in the growth theory of Kuroda and Lacmann \cite{kuroda82}. However, other evidence points to somewhat larger step free energies in the crystal grown from vapor, since the roughening temperature of prism planes grown from vapor occurs at a significantly higher temperature of ca. -2~$^{\circ}$C \cite{furukawa87,elbaum91,gonda99,asakawa15} (c.f. compared to $-16$~$^{\circ}$C for prisms grown from water). In fact, on theoretical grounds such a situation is expected to occur whenever the premelting film  remains of finite thickness as the triple point is approached \cite{benet16,benet19}. So the crucial question is then, is there complete surface melting at the ice/vapor surface (divergence of the premelting film thickness at the triple point)? or is it just incomplete surface melting (finite film thickness at the triple point)?

Despite of discussions and inconclusive results \cite{slater19,nagata19}, there exists a very simple and compelling evidence that not a single plane of ice exhibits complete surface melting \cite{nozieres92,benet19,luengo22b}. The reason is the presence of binding van der Waals forces, which conspire against the unbound growth of the premelting layer. Indeed, 
the effective potential felt by a premelting layer of thickness $h$ at long range in absence of dissolved ions is solely due to van der Waals forces  and is given exactly as \cite{parsegian05}:
\begin{equation}
   g(h) = -\frac{A}{12\pi h^2}
\end{equation}
where $A$ is the Hamaker constant. Based on wetting theory for simple pair potentials  \cite{dietrich91,schick90,evans19,luengo22b}, the sign of the Hamaker constant, $A$, is dictated by the relation $A\propto (\rho_v-\rho_l)(\rho_s-\rho_l)$, where $\rho_{\alpha}$ is the number density of
ice, water or water vapor ($\alpha={s,l,v}$), respectively. Whence, as one yet more anomaly of water, the fact that $\rho_l > \rho_s$ implies that, contrary to premelted films in other solids \cite{chernov88}, the premelting film thickness on ice is always attracted towards the solid surface for large $h$ values, and is therefore not allowed to grow unbound. A positive sign of the Hamaker constant for this system is also obtained from a more sophisticated quantum-electrodynamic account of dipole fluctuations   \cite{elbaum91b,fiedler20,luengo21,luengo22b} based on Dzyaloshinski-Landau-Pitaevskii theory \cite{dzyaloshinskii61}.

Clear experimental evidence for the absence of complete surface melting of ice is the observation of liquid droplets formed on its surface, firstly reported as a testimony of observations \cite{knight67,ketcham69,knight71} but then gradually recorded with images of increasing quality \cite{elbaum91,elbaum93,gonda99,sazaki12,murata16,asakawa16}. Particularly interesting in this regard is the observation of water droplets formed on the surface of fully developed ice crystals at the melting point, which showed that the basal plane remained faceted and incompletely wet as the the prism face melted \cite{gonda99}.

Acknowledging that ice/vapor surfaces exhibit only incomplete surface melting, it must be also acknowledged that the basal plane, which remains smooth up to the triple point, can potentially exhibit some extent of overheating. Being faceted, it is expected that melting will occur in layer wise fashion, with a finite activation energy for the formation of steps \cite{nada05,rozmanov12,neshyba16,hudait16,mochizuki23,cui23}. But a second and less well recognized source of overheating is the mere presence of a premelting film of finite equilibrium thickness. Indeed, the only way the ice surface could melt is by increasing the amount of premelted liquid, which, by definition of equilibrium  is found at a minimum of the surface free energy. Whence, melting requires moving the film thickness away from its equilibrium value over a free energy barrier.

Some hints of the extent of overheating may be found in studies of the TIP4P/Ice model \cite{abascal05}, one of Abascal's important contributions in the field of empirical force fields. On inspection of the literature for melting points obtained from direct coexistence
\cite{vega06,fernandez06,conde17,bore22,ji24}, results appear to fall into two different categories(c.f. Table 1). Those obtained from simulations of the ice/water interface consistently lay just below 270~K \cite{fernandez06,conde17,bore22,addula22,ji24}; those performed from simulations of ice in vacuo suggest a melting point in the range  271-272~K instead \cite{vega06}. The latter estimate appears consistent with  previous work on ice premelting, including our own \cite{llombart19,llombart20,llombart20b},
which have consistently reported equilibrium premelting thicknesses up to 270~K  or 271~K for both prism and basal faces \cite{conde08,kling18}.

Understanding the problem of overheating is interesting {\em per se}, as a fundamental problem in crystal growth, but is also important in order to exercise caution in the determination of melting points by direct coexistence methods. An important feature of
ice is the very steep slope of the melting line, which increases by more than 100~atm per Kelvin. In calculations of coexistence pressure, an error of 1~K therefore can result in changes of pressure estimates that are huge and can impact a number of relevant thermodynamic calculations, such as the Laplace pressure inside a crystal nucleus, or the
interface free energy of a premelting film.


The aim of this paper is to elucidate the extent of ice overheating at the ice/vapor interface by means of computer simulations. Such an endeavor has two requisites. First, a suitable force field for water, and second,  knowledge of its phase boundaries. Abascal and Rull pioneered the art of  molecular simulation in Spain \cite{abascal85,rull85}, and together with Lago and Enciso participated in major efforts for the search of useful intermolecular potentials and the determination of phase boundaries at first order phase transitions \cite{vega03,sanz04,sanz04b,macdowell04b,vega05,demiguel90,demiguel91b,demiguel91c,garzon94,garzon95,romero02,tildesley83,almarza90,almarza92b}.  This paper therefore hinges on a number of topics very well suited to a special issue in their honor

In the next section we discuss how to calculate the interface potential that governs the free energy of premelting films, and how small changes in the estimated melting temperature can  affect its calculation and qualitative behavior. Section III describes the methodology employed in our work, and section IV is devoted to the presentation of results. The most important findings  are summarized and discussed in section V, together with the conclusions.

\begin{table}[h!]
	\centering
	\caption{Literature values for the normal melting point of TIP4P/Ice as obtained from direct coexistence simulations.}
	\begin{tabular}{cclc}
		\hline
		\hline
		Interface  & Time Scale (ns)  & $T_m$ (K) & Reference  \\
		\hline
		\multirow{4}{0.15\textwidth}{ice-water}  & 10       & $268 \pm 2$        & \cite{fernandez06}  \\ 
		  & 100      & $269.8 \pm 0.1$    & \cite{conde17}      \\ 
		  & 300      &  $268.78 \pm 0.1$  &  \cite{ji24}        \\ 
		  & 10       & $269.1$            & \cite{bore22}       \\ 
		  \hline
			\multirow{2}{0.15\textwidth}{ice-vapor}   & 10       &  $271 \pm 1$       & \cite{vega06}       \\ 
		  & 35       & $\approx 271$ & \cite{llombart19,llombart20b,sibley21} \\ 
		\hline
		\hline
	\end{tabular}
	\label{tab:melting-points}
\end{table}


\section{Calculation of interface potentials}

\label{sec:ifp}

The interface potential of a wetting film is a measure of the film's free energy. In practice, it can be viewed as an effective potential for the film thickness, $h$. Systems that exhibit incomplete wetting feature an interface potential with (at least) one absolute minimum at finite thickness, while the interface potential of a surface exhibiting complete wetting has an absolute minimum at $h\to \infty$.

In the theory of wetting, the interface potential is best described in terms of the   surface grand potential, $\omega(h)$  of a system at constant temperature and chemical potential, according to \cite{schick90,dietrich88,henderson05}:
\begin{equation}\label{eq:omegaiv}
  \omega(h) = \gamma_{sl} + \gamma_{lv} + g(h) - \Delta p h
\end{equation}
where $\gamma_{\alpha\beta}$ are surface tensions between phases $\alpha$ and $\beta$, $s$, $l$ and $v$ stand for solid, liquid and vapor phases, respectively; and $\Delta p = p_{l}(\mu,T) - p_{v}(\mu,T)$ is the Laplace pressure difference between the bulk liquid and vapor at the same chemical potential and temperature.

In this equation, $g(h)$ is defined such that it vanishes at $h\to\infty$. Whence, the first two terms account for the surface free energy of a complete wetting film at bulk liquid-vapor coexistence, where $\Delta p=0$. Away from coexistence, the third term favors the phase that has larger pressure at the imposed chemical potential, whence, at supersaturation, with $p_l>p_v$, the growth of a completely wetting film is favored by the bulk contribution; while at sub-saturation, with $p_v>p_l$, the growth of the wetting film is penalized 

Overall, the equilibrium film thickness is dictated by the condition:
\begin{equation}\label{eq:eqcon}
 \left . \frac{d\omega}{dh} \right |_{T,\mu}= 0
\end{equation}
This corresponds to the result:
\begin{equation}\label{eq:calcdisj}
   \Pi(h) = -\Delta p
\end{equation}
where $\Pi(h)$ is the disjoining pressure \cite{derjaguin92b}, defined as
\begin{eqnarray}\label{eq:disjoining}
 \Pi(h) = - \frac{d g(h)}{dh}
\end{eqnarray}

For the special case of premelting, the bulk solid and vapor phases sandwiching the film are all three made of the same substance, so that the equilibrium condition Eq.(\ref{eq:eqcon}) can only hold exactly along the sublimation line. Therefore, in order to calculate the disjoining pressure from Eq.(\ref{eq:calcdisj}), it is required to know the Laplace pressure difference $\Delta p$ along the sublimation line \cite{llombart20,sibley21,luengo22b}.

A simple prescription for the pressure difference between a bulk liquid and its vapor may be obtained from the Gibbs-Duhem equation $dp = \rho d\mu$, by integrating from liquid-vapor to solid-vapor coexistence, under the assumption that the liquid phase has constant density, and the the vapor density is negligibly small. This leads readily to the well know result:
\begin{equation}\label{eq:laplace_route1}
 \Delta p|_{sv}(T) = RT \rho_l \ln \frac{p_{sv}(T)}{p_{lv}(T)}
\end{equation}
where $\Delta p|_{sv}(T)=p_l(T,\mu_{sv}(T))-p_v(T,\mu_{sv}(T))$ is the Laplace pressure between liquid and vapor phases for states $(T,\mu_{sv})$  corresponding to the sublimation line, while
$p_{sv}$ and $p_{lv}$ are  vapor pressures along the sublimation and condensation lines, respectively. The familiar expression most often used to estimate the Laplace pressure in premelting films may be obtained by replacing the Clausius-Clapeyron equations for the coexistence lines into Eq.(\ref{eq:laplace_route1}). This yields: 
\begin{equation}\label{eq:laplace_route2}
   \Delta p|_{sv}(T) = \rho_l \Delta H_t \left( \frac{T-T_t}{T_t} \right)
\end{equation}
where $ \Delta H_t$ is the enthalpy of melting and the subindex $t$ stands for properties pertaining to the triple point. 

As far as the calculation of the disjoining pressure is concerned, the result
serves to highlight that errors in the determination of the triple point  not
only change estimates of the disjoining pressure by an amount $\rho_l\Delta H_t/T_t$,
ca. 10~atm per Kelvin, but also lead to a wrong sign of the disjoining pressure for film heights close to the triple point. 

On quantitative grounds, the result of Equation~(\ref{eq:laplace_route2}) may be improved by integrating the Clausius equation with account of the temperature dependent enthalpy of melting, which leads to:\cite{llombart20,sibley21}
\begin{equation}\label{eq:pvapgen}
\ln (p_{\alpha v}(T)/p_t )=
\frac{ \Delta H_t - \Delta C_{p,t} T_t}{R}
\left [ 1/T_t - 1/T \right ] + \frac{\Delta C_{p,t}}{R}\ln (T/T_t),
\end{equation}
This result may be used in Equation~(\ref{eq:laplace_route1}) and extends the accuracy of Equation~(\ref{eq:laplace_route2}) significantly.

In the next section, we will revisit the calculation of the
triple point of TIP4P/Ice. 
In the remaining part of the section, we will discuss
how to improve on the  results of Eq.(\ref{eq:laplace_route1}), by means of
a suitable thermodynamic integration path.   The calculation is particularly appropriate for this
special issue, as one of Abascal's important contributions to the study of water's phase
diagram was the practical implementation of Gibbs-Duhem and Hamiltonian Gibbs-Duhem integration techniques using a combination of differential equation solvers,  Computer Simulations and batch scripting \cite{sanz04,sanz04b,sanz05}. 

In principle, the problem of finding $\Delta p|_{sv}(T)$ can be solved by integration
of the following exact result \cite{sibley21}:
\begin{equation}\label{eq:long_dp}
\left . d \frac{\Delta p}{dT}\right |_{sv} = \left . \frac{
	\rho_s\rho_l S_l-
	\rho_v\rho_l S_l+
	\rho_l\rho_v S_v-
	\rho_l\rho_s S_s+
	\rho_s\rho_v S_s-
	\rho_s\rho_v S_v}{\rho_s - \rho_v } \right . dT
\end{equation}
where $S_{\alpha}$ stand for molar entropies of phase $\alpha$ along a path $(T,\mu_{sv}(T))$, corresponding to states where the chemical potential is set to its
value at solid-vapor coexistence. This implies that the solid and vapor phase properties are evaluated at the sublimation pressure, while those of the liquid phase are evaluated at a pressure corresponding to the liquid at the same temperature and chemical potential as vapor and solid phases. In practice, the vapor density along the sublimation line is negligibly small compared to that of liquid and solid phases, and the above exact result can be accurately approximated to: 
\begin{equation}\label{eq:laplacesv}
\left . d \frac{\Delta p}{dT}\right |_{sv} = \rho_l \Delta S
\end{equation}
where $\Delta S=S_l-S_s$ is the difference of entropy between liquid and solid phases at the solid-vapor coexistence chemical potential. Notice this corresponds to the entropy of phase change only at $T=T_t$. At other temperatures, we must take into account the entropy increments due to changes in temperature and pressure, according to:
\begin{equation}
dS=\left(\frac{\partial H}{\partial T}\right)_p \frac{dT}{T}-\left(\frac{\partial V}{\partial T}\right)_T dp
\end{equation}
which applies separately for both the liquid and solid phases. Since the path of integration is along states of solid-vapor coexistence $(T,\mu_{sv}(T))$,  the pressure of the solid phase moves along the line of sublimation pressure, but the pressure of the liquid departs strongly from the sublimation line and can increase by hundreds of atmospheres. Therefore, the entropy increments of the solid phase due to changes in pressure is negligible, but that of the liquid phase can become significant. In view of this, we  write:
\begin{equation}\label{eq:entropysv}
 d\Delta S = \Delta C_p \frac{dT}{T} - V_l\alpha_l d p_l 
\end{equation}
where $\Delta C_p= C_{p,l} - C_{p,s}$ is the difference of heat capacity between the solid and liquid phases along the path of integration, while $\alpha_l$ and $p_l$ are the thermal expansion coefficient and pressure of the liquid phase, respectively.

An accurate calculation of the Laplace pressure $\Delta p$ therefore requires solving the set of two coupled differential equations Eq.(\ref{eq:laplacesv}) and Eq.(\ref{eq:entropysv}), in a way reminiscent of the Gibbs-Duhem integration technique used by Abascal in some of his most important papers \cite{sanz04,sanz04b,sanz05}.


\section{Model and Methods}


\subsection{The TIP4P/Ice model}

In this work we perform computer simulations of the ice-vapor interface using the
well known TIP4P/Ice model \cite{abascal05}, one of Abascal's important contributions to computer modeling.  

Some time before the development of this model, a general purpose non-linear least square fit developed by Juan Carlos Gil Montoro, a phD student of Abascal at that time, was combined with batches of Monte Carlo simulations via shell scripting to develop a promising force field for linear and branched alkanes \cite{macdowell98b,macdowell00b}.  This idea served as a precedent to the development of a new generation of water force fields where Abascal's contributions where crucial \cite{abascal05,abascal05b}. Together with the advanced scripting techniques and dedicated simulation protocols mastered by Abascal, the secret for the success of the model  was the use of state of the art free energy methods \cite{sanz04,sanz04b,vega05} and  the optimal choice of thermodynamic target properties for the fit. Particularly, it was crucial for the model's success to combine calculation of ice/water phase boundaries, developed about that time \cite{sanz04,sanz04b,vega05b}, with water properties, such as the temperature of maximum density. During the fitting process, it was soon realized that it was not possible to reproduce accurately both the melting line and the temperature of maximum density. So it was decided to fit two different sets of parameters simultaneously, one targeting  the temperature of maximum density, and other the phase boundaries. The plan is easier said than done, because the preparation of rather complex ice structures such as ice III and ice V, with partial ordering of the hydrogen bond network did not allow to measure the phase boundaries in a direct manner by the Hamiltonian Gibbs-Duhem methodology \cite{vega05b}. Fortunately those problems could be solved by extending theoretical tools and computational methodologies developed for ice Ih \cite{buch98,minagawa81}. This allowed to prepare ice III and ice V configurations with partial disorder consistent with experiment \cite{lobban00}, and to evaluate the Pauling entropy of the partially ordered ice phases \cite{macdowell04b}, which are important in deciding the range of stability of ice III used as  target property for TIP4P/Ice. Somewhat surprisingly for some of us, TIP4P/2005, the model targeting the temperature of maximum density \cite{abascal05b}, turned out to be a better model for liquid water, and gained enormous popularity soon after. However, recent times have seen TIP4P/Ice becoming the preferred choice for the simulation of phenomena where the coexistence of ice and water is relevant, and is therefore the choice of model used in this work.

\subsection{Molecular Dynamics Simulations}

All of the work leading to the phase diagram of water and the development of the new generation of TIP4P water models  \cite{abascal05,abascal05b} was carried out with a dedicated Monte Carlo code developed in the group over many years. Abascal, together with his student  Garc{\'i}a Fern{\'a}ndez pioneered the use of the Molecular Dynamics package  GROMACS in the group \cite{fernandez06}, and devoted a great deal of persuasion for the change of computer simulation paradigm at that time. This is not surprising. The development of the group's code had been a major effort carried out for more than ten years. Starting from a Monte Carlo code for the simulation of dumbbells by Peter Monson, the program was later extended to a multipurpose NVT, NpT code for the simulation of interaction site models of arbitrary geometry \cite{macdowell00b}, including alkanes \cite{macdowell01}. Later came the addition of Ewald sums based on a routine by Bresme, the extension to mixtures \cite{largo02}, and the transformation to allow for Einstein Crystal and Hamiltonian Gibbs-Duhem simulations \cite{vega05,sanz06}. After such effort, some of us were reluctant to shift to GROMACS, but Abascal insisted the Molecular Dynamics package was about twice more efficient on our single processor computers. With the advent of multi core work stations soon after, the discussion on the choice of codes was settled, and the Monte Carlo program completely abandoned except for a few exceptional situations where it remains competitive up to to date \cite{macdowell10,aragones11,aragones11b}.

In this work simulations are carried out using Molecular Dynamics.

The simulation scheme involves several steps. In the first one, an ice slab is built by the replication of a suitable unit cell. In order to prepare roughly square surfaces, we build a unit super-cell of size $(2\times a)\times b\times c$, with $a$, $b$ and $c$,  the unit cell parameters of a pseudo-orthorhombic cell of 8 molecules \cite{macdowell10}. A suitable random hydrogen bond network can then be created following the method of Ref.\cite{buch98}, or alternatively, by disordering the structure of ordered Ice XI, using methods described in \cite{rick03,macdowell10,moreira18}.  A  batch of bulk
simulations in the $NpT$ ensemble is then performed to obtain unit cell parameters of ice Ih as a function of temperature. Another slab of ice with suitable hydrogen bond disorder is  placed in the middle of a simulation box and is rescaled to the lattice parameters of the chosen temperature. Two interfaces with the vapor phase are then created by the elongation of the box by about $10$ nm in the $z$ direction. Depending on the orientation of the replicated unit cell, we create basal (0001), pI ($10\overline{1}0$) or pII ($1\bar{2}10$) interfaces. The resulting configurations are finally simulated in the $NVT$ ensemble, after a suitable period of equilibration.

Results for the melting of ice Ih from the equilibrated configurations are obtained in two different ways. In the first case, results are prolongation of our previous computer simulations\cite{llombart19,llombart20} using the GROMACS package \cite{gromacs4}. In the second case, results are obtained from new simulations of the same system using LAMMPS \cite{lammps22}. This serves to double check our results and to illustrate some subtle differences that can arise when attempting to locate phase boundaries using computer simulations.

\begin{table}[h]
\centering
\caption{Box dimensions  of the systems studied for basal,
primary (pI) and secondary (pII) prismatic facets using LAMMPS. Simulations
are performed with GROMACS with essentially the same settings but $L_z=15.00$~nm
(c.f. Ref.\cite{llombart19,llombart20}).}
\begin{tabular}{c c c c c}
\hline
\hline
Orientation & Temperature (K)  & $L_x$ (nm) & $L_y$ (nm)& $L_z$ (nm)  \\
\hline
\multirow{4}{*}{Basal}& 269  & 7.269597 & 6.291138 & 17.382600  \\
                      & 270  & 7.269957 & 6.291451 & 17.382900   \\
                      & 271  & 7.270317 & 6.291765 & 17.383300  \\
                      & 272  & 7.270677 & 6.292078 & 17.383700  \\
\hline
\multirow{4}{*}{pI}   & 269  & 7.269597 & 5.915552 & 17.826300 \\
                      & 270  & 7.269957 & 5.915849 & 17.826700 \\
                      & 271  & 7.270317 & 5.916147 & 17.827100 \\
                      & 272  & 7.270677 & 5.916435 & 17.827400 \\
\hline 
\multirow{4}{*}{pII}  & 269  & 5.915552 & 6.291137 &  19.020150 \\
                      & 270  & 5.915849 & 6.291451 &  19.020550 \\
                      & 271  & 5.916147 & 6.291765 &  19.021050 \\
                      & 272  & 5.916435 & 6.292078 &  19.021450 \\
\hline
\hline
\end{tabular}
\label{tab:sizes-ice-water}
\end{table}

\subsection{GROMACS settings}

Trajectories in MD simulations with GROMACS were calculated with the leap-frog algorithm using a time step $\tau=3$~fs.  Bonds and angles are constraint using the LINCS algorithm.
Initial ice configurations are equilibrated in the NpT ensemble using the Berendsen barostat. Ice-vapor interfaces are simulated in the NVT ensemble and thermostatted using the velocity rescale algorithm with  1~ps relaxation time \cite{bussi07}. Lennard-Jones interactions are truncated at a distance of 9~{\AA}. Electrostatic interactions are evaluated using Particle Mesh Ewald, with real space cutoff of  9~{\AA}.
The reciprocal space term is evaluated over a total of $80\times 64\times 160$  vectors in the $x$, $y$, $z$ reciprocal directions, respectively. The charge structure factors were evaluated with a grid spacing of 1~\AA and  a fourth order interpolation scheme.  The systems consist of a suitable stack of $8\times 8\times 5$ super-cells with a total of N=5120 molecules.

\subsection{LAMMPS settings}

Trajectories in MD simulations with LAMMPS simulation package \cite{lammps22} were evolved using the velocity Verlet integration algorithm with a time step $\tau=2$~fs. 
In the case of the ice-vapor coexistence the temperature was maintained using the
velocity rescaling algorithm with a damping factor equal to $\tau_{VR}=2$ ps \cite{bussi07}.
In the auxiliary $NpT$ simulations, both the temperature and pressure were maintained
using the Nos{\'e}-Hoover chains algorithm  \cite{martyna92, martyna94}  with a damping factors equal to $\tau_{NH}=2$ ps and number of chains equal to 3. 
Lennard-Jones interactions were truncated at a distance $9$ \AA~ and the tail correction
was included. Long-range electrostatic interactions were calculated using
the particle-particle particle-mesh method \cite{hockney88}. 
The charge structure factors were evaluated with the fourth-order interpolation scheme
and a grid spacing of $\approx1$ \AA~resulting in the following number of k-vectors in the reciprocal space for 
(i) basal: 72, 64, 180, (ii) pI: 72, 60, 180, and (iii) pII: 60, 64, 180 faces, 
in the $x, y, z$ directions, respectively (cf. Table~\ref{tab:sizes-ice-water}).
However, notice slight differences in the box sides $L_x$, $L_y$, and $L_z$ depending on the exposed facet of the ice Ih crystal with the vapor phase as shown in Table~\ref{tab:sizes-ice-water}. The systems consist of a suitable stack of $8\times 8\times 10$ super-cells with a total of N=10240 molecules.

\subsection{CHILL+ order parameter}

In order to distinguish the solid-like and liquid-like environments we used the CHILL+ parameter \cite{nguyen15}.
Discrimination was based on the calculation of the $c_3(i,j)$ correlation function 
allowing to identify nearest neighbor molecules in staggered ($c_3(i, j) \leq -0.8$) and eclipsed conformations
($-0.05 \geq c_3(i, j) \geq -0.2$) and discriminate between different allotropic environments. Molecules attributed to ices Ih and Ic, including interfacial Ih molecules, are labeled as solid like, while all other remaining types, including mislabeled molecules, are attributed to the liquid phase.

\subsection{Thermodynamic integration}

The Laplace pressure difference $\Delta p|_{sv}$ required to calculate the disjoining pressure curve is obtained by solving the system of two coupled differential equations Eq.(\ref{eq:laplacesv}) and Eq.(\ref{eq:entropysv}) on the fly,
as in Gibbs-Duhem integration.

At each state $(T',p_l')$, $\Delta p|_{sv}(T)$ is calculated by a simple Euler solver from data collected at $(T,p_l)$, such that:
\begin{equation}\label{eq:solverp}
 \Delta p|_{sv}(T') =  \Delta p|_{sv}(T) + \frac{\Delta S(T)}{V_l(T,p_l)}(T'-T)
\end{equation}
The entropy change required in the equation above is estimated as: 
\begin{equation}\label{eq:solvers}
\Delta S(T')=\Delta S(T)+\Delta C_p(T)\ln\frac{T'}{T}-V_l \alpha_l(p'-p)
\end{equation}
Thermodynamic quantities of the solid phase are calculated at the corresponding temperature, under the assumption that $p_v=p_s=1$~atm, which results in negligible error; while the properties of the liquid phase are calculated at the corresponding temperature, and at an estimated liquid phase pressure of $p_l(T) = \Delta p|_{sv}(T)$.

The thermodynamic integration starts at the assumed triple point, where $T=T_t$, $p_v=p_s=p_l=1$~atm, $\Delta S=\Delta H_t/T_t$ and $\Delta p|_{sv}(T_t)=0$.  For a prescribed temperature $T'=T+\delta T$, we find a new $\Delta p|_{sv}(T')$ using the solver Eq.(\ref{eq:solverp}). We then perform simulations at $T',p_s=1$~atm, for the solid and at $T',p_l$ for the liquid phases to estimate the derivative properties  as:
\begin{align}
C_p(T)=\left(\frac{\partial H}{\partial T}\right)_p&=\frac{H(T+\delta T,p)-H(T,p)}{\delta T} \nonumber \\
V\alpha(T)=\left(\frac{\partial V}{\partial T}\right)_p&=\frac{V(T+\delta T,p)-V(T,p)}{\delta T}
\label{eq:derivatives}
\end{align}
These results are used to  update $\Delta S(T')$  according to the solver Eq.(\ref{eq:solvers}). The process then starts again, with simulations of the solid phase at $T'$ and pressure $p_s=1$~atm, and simulations of the liquid phase at $T'$ and a pressure $p_l= \Delta p|_{sv}(T')$.

Once the full curve $\Delta p|_{sv}(T)$ is known, use of results for the premelting film thickness at solid-vapor coexistence, $h(T)$ obtained in Ref.\cite{llombart20} are used to determine the disjoining pressure $\Pi(h(T))=-\Delta p|_{sv}(T)$ according to Eq.(\ref{eq:calcdisj}).

\section{Results}

 \begin{figure}[htb]
	\centering
	\includegraphics[width=\linewidth]{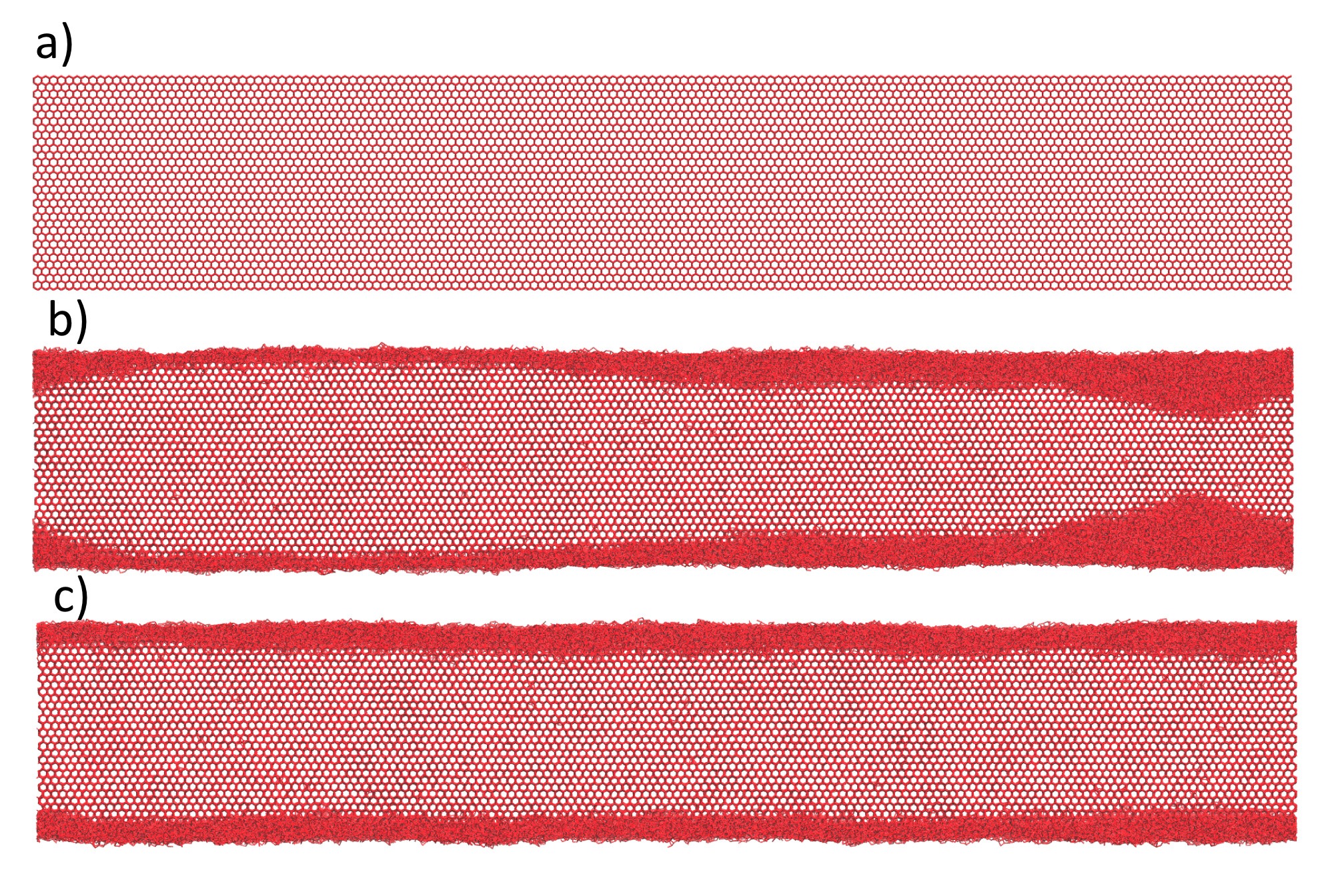}
	\caption{
	\label{fig:incident}
Direct observation of overheating. a) Snapshot of an initial perfect bulk ice slab placed in vacuum with the prism face exposed. b) Long lived metastable state observed for single precision GROMACS simulation with temperature targeted to $T=272$~K. c) State of an initial configuration as drawn from b) after 25~ns of a GROMACS Molecular Dynamics run with double precision.}
 \end{figure}

\subsection{Overheating incident}

Before we describe our results in detail, it is interesting to discuss the outcome of a preliminary simulation we performed for our project 
on the surface structure of the ice-vapor interface some time ago \cite{llombart20b}. At that time, we simulated the prism face  of a very large system with 
box dimensions of $L_x=72.68$, $L_y=11.83$ and $L_z=20.29$~nm containing 327,680 molecules overall. 
Starting from a perfect ice crystal (Figure \ref{fig:incident}-a),  we
performed Molecular Dynamics in the NVT ensemble using GROMACS v.2016.4 at
single precision, aiming at a target temperature of $T=272$~K.  
During the course of the simulation, the system first formed a premelting film, as expected, but subsequently entered into a long lived 
metastable state with a small water droplet that persisted for about 100~ns(Figure \ref{fig:incident}-b). Strikingly, the droplet did not stick out into the bulk vapor
phase, but instead penetrated into the bulk solid with  the water/vapor surface remaining almost flat at the location of
the droplet. Since the curvature of an interface dictates the local excess pressure over bulk \cite{degennes04}, we conclude that: i) the premelting
liquid film exhibited a perpendicular pressure that was very close to that of the vapor phase, i.e., essentially zero pressure and ii) the pressure of the
liquid droplet was larger than that of the bulk solid. 

Since the liquid and solid phases can mutually feed one from the other, they must be
found at equal chemical potential. A thermodynamic state where liquid and solid phases have equal chemical potential, but the liquid phase has larger pressure 
than the solid is actually a thermodynamic state where the liquid is the stable phase and the solid is metastable: Whence, our system must have been in a state of overheating. In fact, 
because the bulk vapor pressure is close to zero and $p_l>p_s$,  the pressure of the bulk solid was likely negative, 
and the solid must have   been found under tension. This corresponds to an overheated  state with $T>T_m$, but at pressures below the liquid-vapor
coexistence pressure, such that the vapor is more stable than the liquid. 

As another interesting observation, we note that the premelted film that was formed appeared to display also two different film heights. Curiously, such a state can
be viewed as the mirror image of experimental observations performed on the ice surface close to melting at slight supersaturation. Under those
conditions, it is possible to find transient states with two different quasi-liquid layer thicknesses, coexisting with a liquid droplet that
grows into the vapor phase \cite{asakawa15,murata16}. However, this observation could well be a fortuitous coincidence and should be taken with some reservations.

The follow-up of this anecdote is a word of caution for GROMACS practitioners. On repeating the simulation at the same conditions, we obtained similar states, but
the droplet always grew in the right hand side corner of the simulation box. Suspicious of this situation, we decided to restart simulations of
the state observed in Fig.\ref{fig:incident}-b using double rather than single precision. The heterogeneities of the system then vanished within 25~ns,
and a uniform premelted film was formed Fig.\ref{fig:incident}-c.

At the time we interpreted that our single precision simulations suffered from rounding-off errors in the evaluation of the forces, which, together with a choice
$dt=3$~fs, resulted effectively in the modification of the Hamiltonian \cite{davidchack10,rozmanov12}. Additionally, we speculate that some domain decomposition of the system performed by GROMACS could result
in an effective heating up in the right hand side of the simulation box. Reports of spurious pressure anisotropy resulting from inaccurate neighbor-list updates could also be related with this problem \cite{kim23}. The observation was thus disregarded for further analysis, but has ever since left
us with the suspicion that overheating of ice at the ice-vapor interface could be possible.

\subsection{Determination of melting points}

\begin{figure}[h!]
	\centering
\begin{subfloat}[]
	\centering
	\includegraphics[width=0.47\linewidth]{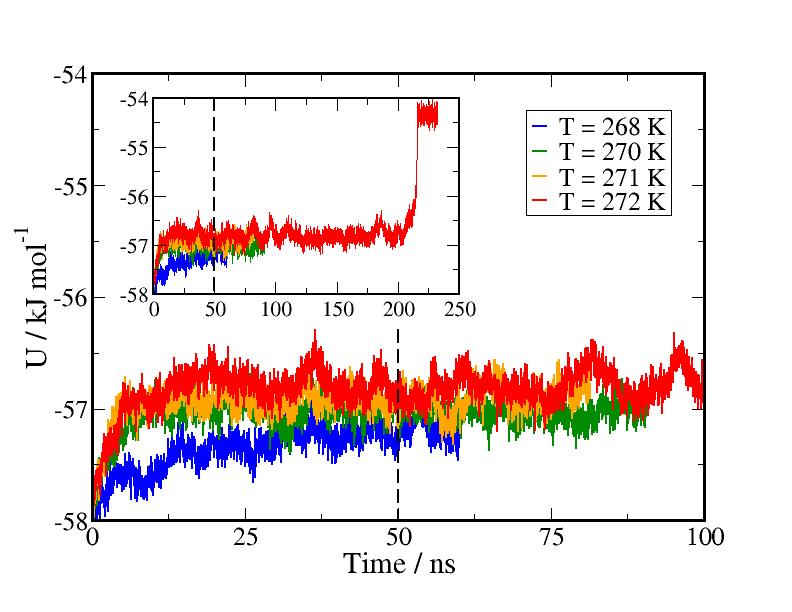}
\end{subfloat}
\begin{subfloat}[]
	\centering
	\includegraphics[width=0.47\linewidth]{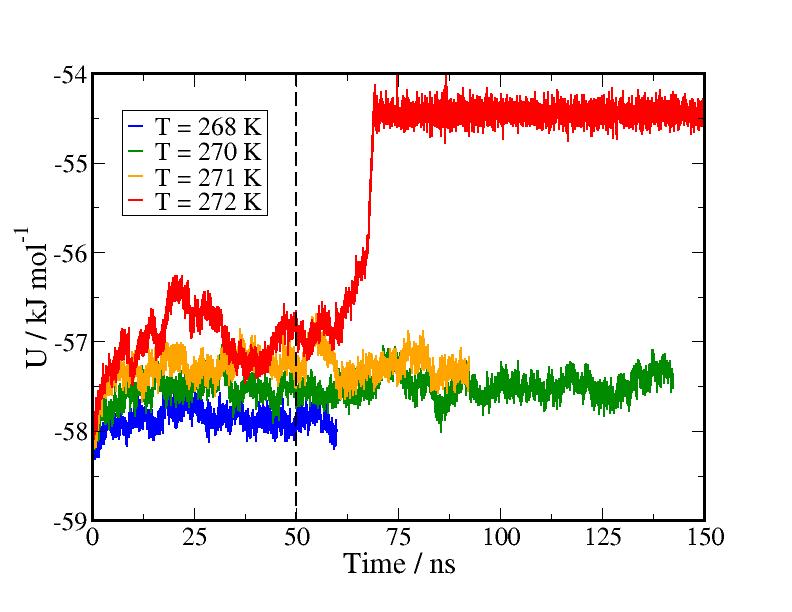}
\end{subfloat}

	\caption{GROMACS simulation of the ice/vapor interface. The figure shows the potential energy as a function of time during the course of simulations of the ice/vapor interface. From top to bottom, the curves show results from T=269 to T=272~K for basal (a) and pI (b) faces. A black dashed line indicates the length of simulations performed in our previous work \cite{llombart20}.}
	\label{fig:melting-faces-gromacs}
\end{figure}

\begin{figure}[h!]
    \centering
    \begin{subfloat}[]
    	\centering
    	\includegraphics[width=0.49\linewidth,trim= 0 0 3cm 0]{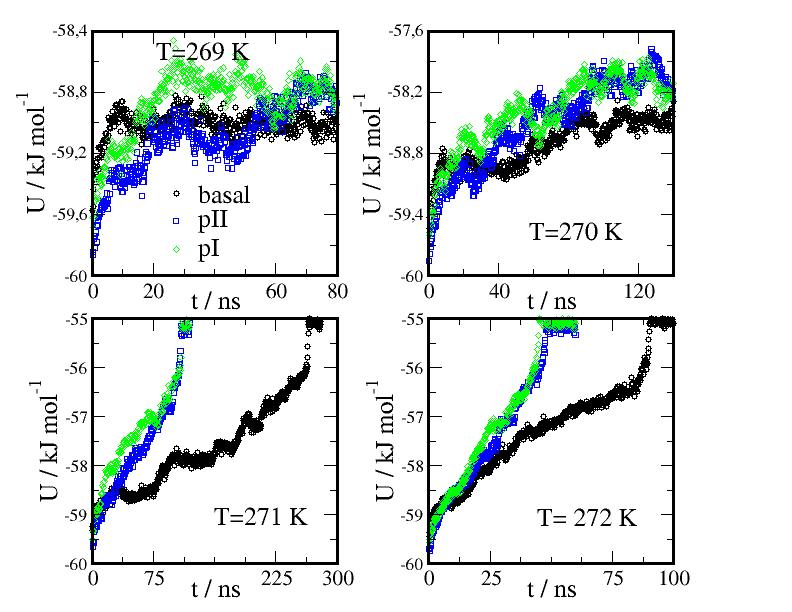}
   	\end{subfloat}
   	\begin{subfloat}[]
   		\centering
   		\includegraphics[width=0.49\linewidth,trim= 0 0 3cm 0]{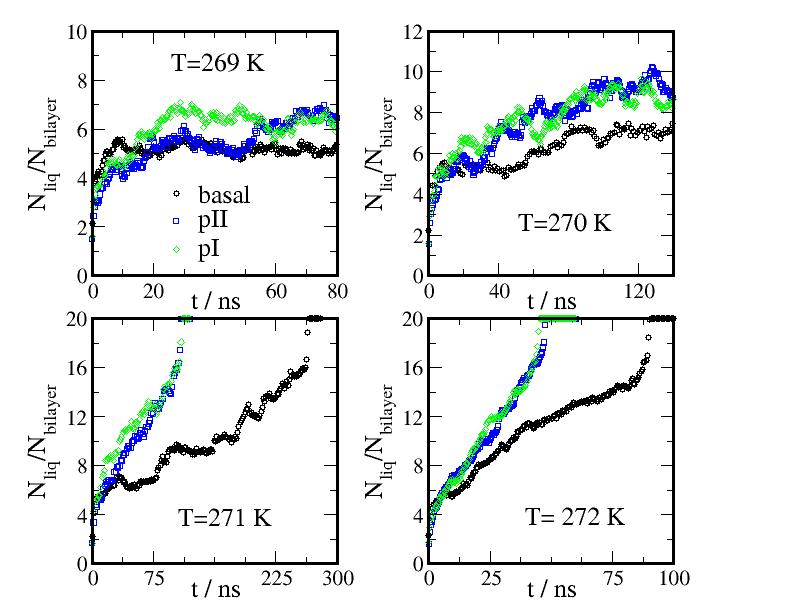}
   	\end{subfloat}
    \caption{LAMMPS simulation of the ice/vapor interface. The figure shows (a) the time evolution of potential energy 
    and (b) fraction of  liquid molecules relative to the total amount of molecules in one single bilayer. Results are shown   for temperatures in the range of $T=269-272$ K.}
    \label{fig:melting-faces-lammps}
\end{figure}

In view of the above observation and the discrepancies for the melting point as shown in Table~\ref{tab:melting-points},
we decided to address this issue and perform new MD simulations of ice in coexistence with its vapor.

Firstly, we report results from  our previous work on ice premelting using GROMACS \cite{llombart19,llombart20}. 

Figure Ref.(\ref{fig:melting-faces-gromacs}) displays the evolution of the potential energy during the simulation of the ice-vapor interface for both basal and pI faces. The results for premelting thickness in our previous work were obtained from averages of the last 35~ns of 50 ~ns simulations. The figures show that, within that time span, both the basal and pI surfaces appear stabilized up to a temperature of 272~K. However, the highest temperature studied, T=272~K, does show rather large energy fluctuations. For that reason, our previous work only reported premelting thicknesses for temperatures up to 271~K as a measure of caution. 

In view of recent reports of a melting temperature of about 269~K for the TIP4P/Ice model, we decided to extend those simulations well beyond 100~ns (c.f. Table I). In this case, we observe significant differences between the basal and pI faces. Particularly, the simulations for the basal face appear to remain stable up to 100~ns for temperatures below 271~K. More strikingly, the  simulation performed at T=272~K, which are 2~K above estimated melting from direct coexistence of the solid/liquid interface remained stable up to the end of our 140~ns simulation.

A similar stability was observed for the simulations in the range 268-271~K for the pI face, but the simulation at 272~K showed clear evidence of melting as revealed by a sharp increase of the potential energy at about 60~ns.

The conclusion from these observations is that the basal face appears to exhibit a larger stability and hence, a slower melting dynamics than the pI face. However, the result for the pI face implies a melting point below T=272~K. In view of this, we can infer that the pI face can remain overheated for about 60~ns, while the basal face can exhibit overheating for time scales as large as 140~ns, which is well above the length of many direct-coexistence simulations.   

This observation is consistent with a scenario where the basal face remains smooth above the melting point \cite{benet16}. In such situations, growth becomes an activated process. This means the crystal surface remains mechanically stable over periods smaller than the nucleation time. Accordingly, no melt/growth appears to occur at all. Whence, the determination of melting points from direct coexistence becomes  unreliable. This is opposite   to the situation found for rough interfaces, where growth or melting will be detected within limited simulation times as the occurrence of growth curves with, possibly small, but finite slope at all temperatures away from coexistence, as expected from a Wilson-Frenkel law.

Unfortunately, the batch of GROMACS simulations does not settle the issue on the melting temperature of the TIP4P/Ice model, since all the simulations appear to exhibit a stable (or at least metastable) solid phase up to 271~K, whereas direct coexistence simulations of the ice/water interface suggest melting temperatures below 270~K \cite{fernandez06,conde17,bore22,addula22,ji24}.

Our results could still be consistent with reports of melting below 270~K, under the understanding that at the ice/vapor interface the basal and pI faces at T=271~K remain metastable over time scales beyond 100~ns. However, we do note that our simulations using GROMACS were performed for a time step $\delta t=3$~fs, which, in retrospective, we realize is somewhat too high \cite{rozmanov12,ashtagiri24}. Because of the use of a robust integrator, the simulations are known to remain consistent with Hamiltonian dynamics for large time steps, but the actual Hamiltonian that is simulated is changed by terms of order $\delta t^2$. Accordingly, our results could  suffer from discretization errors in the determination of the melting point \cite{davidchack10}. Furthermore, the system size might also be somewhat small in the $z$ direction, and reports have been given of an increase in melting temperature for small system sizes \cite{rozmanov11}.

In order to further explore these issues, we have performed a full new set of simulations with similar settings, but with a  system twice as large ($N=10240$ molecules) and a smaller time step $\delta t=2$~fs, using LAMMPS. However, noticing that the alternative pII plane  has been reported to exhibit the fastest melting dynamics, we have also performed simulations with solid/vapor interfaces exposed at that plane  \cite{nada05,rozmanov12}.

The evolution of the potential energy (a) and number of liquid molecules (b) with time for basal, pI and pII faces is  displayed in Figure~\ref{fig:melting-faces-lammps}.

 At $T=269$~K, it is evident that the systems only melt to a limited degree,  reaching equilibrium premelting film thickness in the time scale of 30~ns for all three planes. At $T=270$~K, both energy and number of liquid molecules grow for as long as 100~ns, but basal and pI faces appear to stabilize for the next 50~ns. The simulation time required for the premelting film to equilibrate are similar for all three faces. On the other hand, at the temperature $T=271$ K, that is above the estimated melting point of TIP4P/ice water model, the differences start to clearly emerge. All three systems do eventually melt completely, but the simulation time required  is almost three times longer for the basal facet compared to the prismatic planes. Particularly, the pI and pII faces melt
 at a very similar steady rate, leading to complete liquefaction after 100~ns. However, the basal face appears to grow a liquid layer by sudden and discontinuous jumps, with  apparently equilibrated plateaus lasting about 50~ns in between two such events. Henceforth, the system enters an apparently steady melting regime which takes up to 300~ns for completion.

Hints of the (bi)layer-by-(bi)layer melting have been also  reported for direct coexistence simulations of ice grown from the melt \cite{nada05,rozmanov12,mochizuki23} and the vapor, \cite{neshyba16,hudait16,cui23}. In our work, this is revealed as a step-wise increase of either the potential energy--Figure~(\ref{fig:melting-faces-lammps}-a) or the number of liquid-like molecules--Figure~(\ref{fig:melting-faces-lammps}-b). The existence of such plateau regions support the claim of a smooth basal surface with activated growth, and, accordingly, the possibility of overheating within the scale of a typical molecular simulation.

As regards the melting point, the results from this batch of simulations, with larger system sizes and smaller time step, suggest a melting point that is above 269~K, but below 271~K. This clearly appears a lower melting point than that inferred from the GROMACS simulations, and brings the result in  agreement with estimates obtained from the direct coexistence of ice with water, which point to a melting point slightly below 270~K. Clearly, small differences in the simulation protocol, such as choice of time step and system size can alter the melting point. However, it is not possible to tell whether the relative stability that we find at 270~K is a matter of slower dynamics of the ice/vapor interface, or a purely technical issue related with the simulation protocol.

\subsection{Interface potential}

\begin{figure}[h!]
	\centering
	\includegraphics[width=\linewidth]{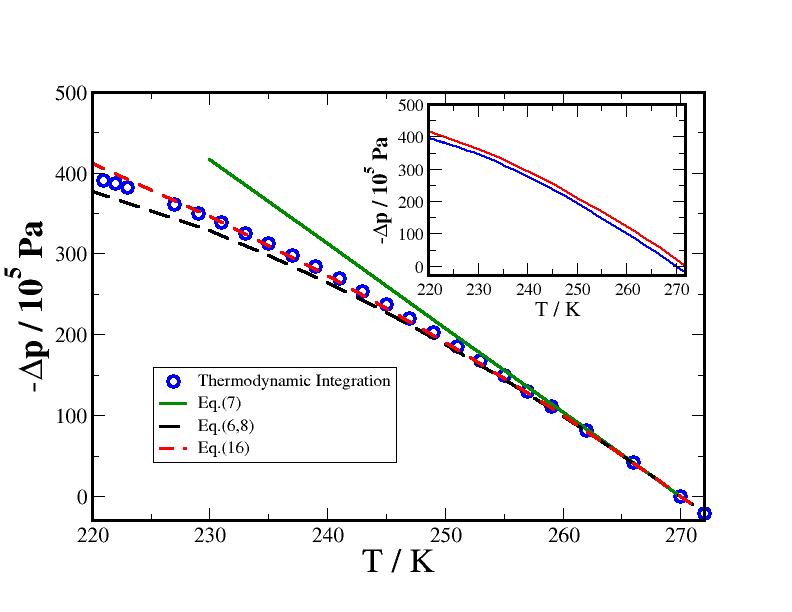}
	\caption{Laplace pressure of premelting films. The symbols are a quadratic fit to the thermodynamic integration data of Table \ref{disjoin}. The full line is the linear approximation of Eq.(\ref{eq:laplace_route2}). The long dashed line is obtained from the combination of Eq.\ref{eq:laplace_route1} and Eq.(\ref{eq:pvapgen}) and the short dashed line is Eq.(\ref{eq:laplace_r5}).    The data required for these models is provided in Table \ref{therm}. The inset compares the thermodynamic integration of this work with results from Ref.\cite{llombart20} assuming $T_t=272$~K.
	\label{fig:disj_v_T}}
\end{figure}

Having settled the issue concerning the melting temperature of TIP4P/Ice, we are now in position to reassess our  previous estimates for the disjoining pressure of premelting films \cite{llombart20,sibley21}.

As described in Section \ref{sec:ifp}, this first requires to estimate the Laplace pressure, that
we have calculated from the thermodynamic integration path Eq.(\ref{eq:laplace_route2}). 
The thermodynamic path starts at the model's triple point, which we have assumed equal to
the normal melting temperature, whence, at $T=270$~K. With negligible error, we have
further considered the triple point pressure at $p=1$~atm, and that ice remains at this pressure along the sublimation line.  The values of melting entropy $\Delta S_m$ and water's molar volume $V_m$ at coexistence required at the start of the integration path are shown in Table~\ref{therm}, as calculated  for a cutoff equal to  $9$ \AA. The results differ slightly from  Table IV in Ref.~\cite{abascal05}, likely due to the differences in cutoff distance and assumed lower value of the melting temperature. 

Our results for the integration path with finite temperature increments of 10~K are collected in Table~\ref{disjoin}. A quadratic fit to the data is displayed in Figure~(\ref{fig:disj_v_T}), together with a comparison of previous estimates of the Laplace pressure as obtained under the assumption that the triple point of TIP4P/Ice was found at $T_t=272$~K \cite{llombart20,sibley21}.

The improved results show a small difference with our previous estimate \cite{llombart20,sibley21},
which has now decreased by  a constant offset of about 20~atm (c.f. Inset of Fig.(\ref{disjoin})).

 However, the significant difference is that we can now confirm the observation of premelting films with negative disjoining pressure.  Indeed, by definition the disjoining pressure is positive below
the triple point (c.f. Eq.(\ref{eq:laplace_route1})). Therefore, our previous results assuming $T_m=272$~K provided a positive disjoining pressure of ca. 20~atm at 270~K. The new results provide by construction null disjoining pressure at this temperature, and a negative disjoining pressure of -10~atm at 271~K. Since the disjoining pressure is minus the slope of the interface potential (c.f. Eq.(\ref{eq:disjoining})), our results now confirm directly by computer simulations the existence of a minimum of $g(h)$, as predicted previously on theoretical grounds \cite{elbaum91b,sibley21,luengo22b}. The results also show explicitly the possibility of overheating ice by about 1~K in typical simulation time scales which are often a few decades of nanoseconds.

Asides this qualitative conclusion, Fig.(\ref{disjoin}) compares the results for $\Delta p$ as obtained by thermodynamic integration, with a number of theoretical models of increasing sophistication. 

The linear approximation of Eq.(\ref{eq:laplace_route2}), which is often used in studies of nucleation performs very well down to -10~C,  but deteriorates significantly as the temperature decreases. To put numbers, we find  errors increasings gradually from 5~\% at $T-T_t=-10^{\circ}$C to 10~\% at $T-T_t=-20^{\circ}$C and 20~\% at $T-T_t=-30^{\circ}$C.
Notice such errors could produce extremely large effect in the estimation of  nucleation barriers, since the Laplace pressure difference for ice nucleation in water at atmospheric pressure is essentially equal to that calculated here. On the other hand, the ideal gas model of Eq.\ref{eq:laplace_route1}, with improved input for coexistence values from a second order Clausius-Clapeyron approximation, Eq.(\ref{eq:laplace_route2}),  remains quantitatively accurate down to 250~K, and qualitatively captures the bending of $\Delta p$ down to 220~K. 

Finally, we have also tested a model based on the direct integration of Eq.(\ref{eq:laplacesv}) using a first order Taylor expansion in the temperature for $\rho_l$ and $\Delta S$. This provides:
\begin{equation}\label{eq:laplace_r5}
 \Delta p|_{sv} = \rho_l \frac{\Delta H_m}{T_t}\Delta T + \frac{1}{2}\left[\rho_l \frac{\Delta C_p}{T_t} - \rho_l\alpha_l \frac{\Delta H_m}{T_t} \right] \Delta T^2 - \frac{1}{3}\rho_l\alpha_l\frac{\Delta C_p}{T_t} \Delta T^3
\end{equation}
Somewhat unexpectedly, this apparently less sophisticated approach works very well in the whole temperature range from 270 to 220~K explored in our simulations.

\begin{table}
\caption{Thermophysical properties of TIP4P/Ice model required for the determination of the
water pressure along the line of equal chemical potentials of the bulk (ice) and 
adsorbed (water) phases at melting temperature $T_m=270$ K under ambient pressure $p=1$ atm.
Additionally, we calculated $V_l\alpha_l=-1.08\times 10^{-8}$~m$^3$K$^{-1}$mol$^{-1}$ and $\Delta C_p=40.69$~JK$^{-1}$mol$^{-1}$ for use in Eq.(\ref{eq:laplace_r5})}
\centering
\begin{tabular} {ccccc}
\hline
\hline
$V_w$ (m$^3\cdot$ mol$^{-1}$) & $\Delta H_m$ (J$\cdot$ mol$^{-1}$)  &$\Delta S_m$ (J$\cdot$ K$^{-1}\cdot$ mol$^{-1}$) & $T_m$ (K) & $p_i$ (atm)\\
\hline
$1.82\cdot10^{-5}$ &$5119.848$ & $18.962$ & 270 & 1 
\end{tabular}
\label{therm}
\end{table}

\begin{table}
	\centering
	\begin{tabular} {ccccc}
		\hline
		\hline
		$T$ (K) & $V_w\cdot 10^5$ (m$^3\cdot$ mol$^{-1}$)  & $p_w$ (atm) & $\Pi^{(1)}$ (atm) & $\Pi^{(2)}$ (atm) \\
		\hline
		230 &        & -343.8 & 344.8   & 361.5     \\
		240 & $1.92$ & -274.4 & 275.4   & 294.0     \\ 
		250 & $1.88$ & -194.2 & 195.2   & 210.1    \\
		260 & $1.85$ & -101.6 & 102.6   & 123.4    \\
		270 & $1.82$ &  1     &  0      &  21.6   \\
	\end{tabular}
	\caption{Results for the thermodynamic integration path   described in Section~\ref{sec:ifp} and Eq.(\ref{eq:solverp}-\ref{eq:derivatives}). Water's volume, pressure and disjoining pressure along the path are estimated at different temperatures. Disjoining pressures are given as calculated from (1) the integration path (2) using Eq.(\ref{eq:laplace_route1}) and (\ref{eq:pvapgen}) with assumed melting point $T_m=272$~K, as in \cite{llombart20,sibley21}.
	}
	\label{disjoin}
\end{table}

\begin{figure}
	\includegraphics[width=0.5\linewidth]{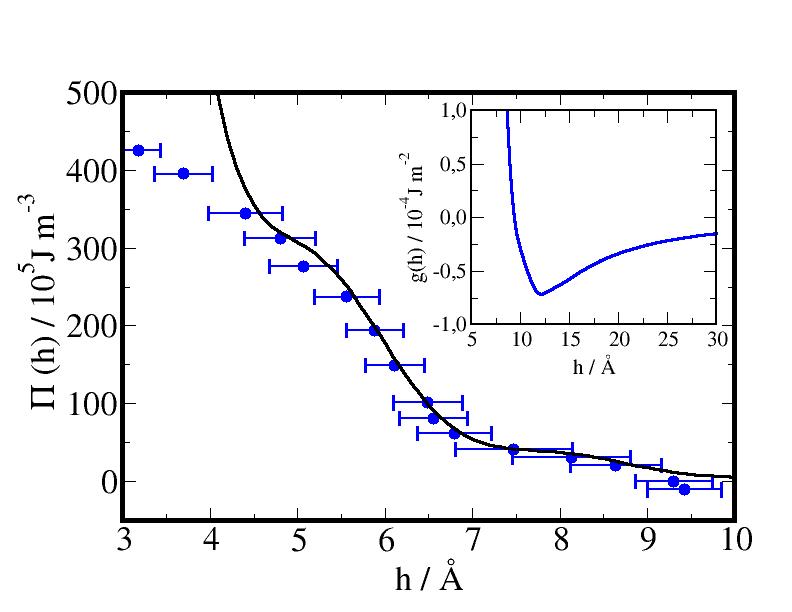}
	\includegraphics[width=0.5\linewidth]{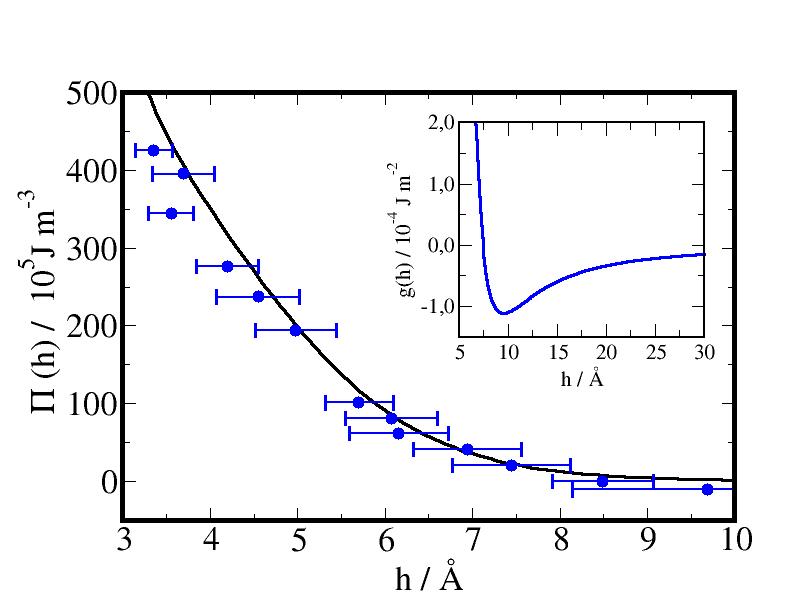}
	\caption{Disjoining pressure of premelting films. Figure shows the disjoining pressure of premelting films formed on basal (left) and prism faces. Symbols represent film thicknesses measured previously \cite{llombart20}, matched to Laplace pressures obtained in this work from thermodynamic integration. The lines are a fit to the results using the model of Eq.(\ref{eq:ifpm}). Insets display the model interface potentials resulting from the parametric fit to disjoining pressures.
		\label{fig:disjoining}}
\end{figure}

Having calculated $\Delta p$, we can now exploit previous results for the premelting film thickness \cite{llombart20}, in order to map $\Delta p(T)$ into $\Pi(h)$ as described in Eq.(\ref{eq:eqcon}). The results for the disjoining pressures of premelting films at the basal and pI planes are displayed in Fig.(\ref{fig:disjoining}). Clearly, the scale of pressures and film thicknesses are broadly the same for both facets, but the detailed shapes are non-universal and depend much on the choice of crystal plane (c.f. Ref.\cite{limmer14}).

Using results from the statistical mechanics of interfaces, the interface potential of premelting films may be decomposed into two contributions \cite{derjaguin87}:
\begin{equation}\label{eq:ifpm}
  g(h) = g_{sr}(h) + g_{lr}(h)
\end{equation}
where $g_{sr}(h)$ is a short range exponential term due mainly to packing correlations, and $g_{lr}$ is an algebraic long range term due to dispersion interactions. 

According to liquid state theory, the short range contribution is expected to obey the following asymptotic form \cite{chernov88,henderson94,evans94}:
\begin{equation}\label{eq:gsr}
g_{sr}(h) =  B_2 \exp(-\kappa_2 h) - B_1 \exp(-\kappa_1 h) \cos(q_o h - \phi)
\end{equation}
where $B_i$ and $B_2$ are positive constants that set the amplitude of the correlations,
$\kappa_1$ and $\kappa_2$ describe the range of short range correlations in renormalized fashion \cite{chernov88}, and $q_o$ is a wave-vector dictating the scale of (possible) oscillations.  

In principle the long range contribution stems mainly from electronic dipole fluctuations, which require a rather involved statistical mechanical treatment that is embodied in the Hamaker constant, $A_H$ \cite{dzyaloshinskii61,fiedler20,luengo22b}:
\begin{equation}
 g_{lr}(h) = - \frac{A_H}{12\pi h}
\end{equation}
In practice, for a simple point polarizable model as TIP4P/Ice, all such effects are lumped into the $\epsilon$ and $\sigma$ Lennard-Jones parameters, and can be described qualitatively as \cite{luengo22b}:
\begin{equation}\label{eq:lrc}
 A_H = 4\pi^2 \epsilon \sigma^6 (\rho_s - \rho_l)(\rho_v-\rho_l)
\end{equation}
Using triple point data for TIP4P/Ice \cite{abascal05}, we find  $A_H=5.07\cdot 10^{-21}$~J for this model (this value differs somewhat from estimates in Ref.\cite{luengo22b}, since experimental triple point data were used in that case). Notice that solid and liquid densities differ only by a small amount, so that the Hamaker constant is rather small and the term in Eq.(\ref{eq:lrc}) contributes negligibly to the interface potential and disjoining pressures for $h$ in the scale of 3 to 10~\AA~ measured in the simulations.

In order to model the interface potential, we fit the disjoining pressure resulting from Eq.\ref{eq:ifpm} to the disjoining pressure curve of Fig.(\ref{fig:disjoining}). Unfortunately, the fit is somewhat under constrained because of the large number of unknown parameters in Eq.(\ref{eq:gsr}). For this reason, the least squares method provides several local minima with different parameters and very similar fit quality. Table~\ref{tab:parameters} collects one such choice with parameters that are order of magnitude close to expectations, but have likely  limited physical significance.  Furthermore, we note that the simulation results suffer from truncation of the long range $r^{-6}$ tail of the Lennard-Jones potential, and need not obey Eq.(\ref{eq:lrc}) strictly. For this reason, the simulation results are actually fitted to a model of truncated Lennard-Jones interactions. This part of the interactions is then subtracted from the fit, and replaced by the exact long range model, Eq.(\ref{eq:lrc}), to provide our final estimate of interface potentials displayed in the insets of Fig.(\ref{fig:disjoining}).

In practice, the fit is hardly affected by this complication, and we leave details of the model truncated interactions for discussion in the Appendix.

The figure shows as expected from the disjoining pressure isotherms that the order of magnitude of surface free energies and film thicknesses are rather similar for both basal and prism faces. However, the basal face exhibits a minimum that is shallower and occurs at larger film thicknesses than that of the prism face. This implies that the basal face is actually closer to complete surface melting than the prism face for TIP4P/Ice. At any rate, the surface free energy at the minimum of the interface potential is barely 0.1~mJ/m$^2$, and thus provides an extremely small correction to the surface free energy of the ice-vapor surface that would be estimated under the assumption of complete surface melting, which is $\gamma_{sl}+\gamma_{lv}\approx 105$~mJ/m$^2$ (c.f. Eq.(\ref{eq:omegaiv})). This does not upset the conclusion that premelting films must exhibit incomplete surface melting for ice in water vapor, but it helps explain why any small perturbation (small deviations from strict ice/vapor coexistence or impurities) can change this situation and produce much thicker films \cite{wettlaufer99}. The shallowness of the minima also imply that the long activation times found for the melting of the basal plane are not related to the presence of premelting, but rather, to the smoothness of the basal facet.

\begin{table}[htb!]
	\center
	\begin{tabular}{c c c c c c c}
		\hline
		\hline
		face & B$_2$ /$J\,m^{-2}$ &  B$_1$/$J\,m^{-2}$ & $\kappa_2 \cdot$\AA & $\kappa_1$
		$\cdot$\AA &
		$q_{o}\cdot$\AA  &  $\phi$/rad \\
		\hline
		Basal &  0.169   &  0.0112  &  0.738  &  0.723  & 2.266  &  2.282  \\
		\hline
		pI  &  0.170   &  0.0912  &  0.880  & 1.118  &  0.947  &  2.152  \\
		\hline
		\hline
	\end{tabular}
 \caption{Parameters for the fit of simulated disjoining pressures to the model
 	interface potential in  Eq.(\ref{eq:ifpm}). Results are shown for basal and prism
 	planes. See appendix for details of the fitting function to a truncated Lennard-Jones tail.  }
 \label{tab:parameters}
\end{table}


\section{Conclusions}

In this work we have studied the likelihood of overheating ice.

We notice that in principle a mono-crystal of whatever substance can be overheated, whether immersed in the melt or the vapor phase, provided the exposed surfaces are smooth (in the statistical mechanical sense). In such situation, crystal growth, melting or evaporation are actually activated events. It is therefore possible to stabilize the crystal at temperatures above melting in the time scale of a two dimensional nucleation event. Of course, such time scale will depend on the scale of the activation energy and the extent of overheating, and will usually involve very short times that are difficult to monitor in conventional experiments. 

As regards ice, our study of the TIP4P/Ice model shows that the basal face can remain in mechanical equilibrium above the melting point for times as large as 100~ns, which is a time span still quite large for usual computer simulations. The melting process appears to occur layer-wise, with melting events that occur suddenly over a small period of time, followed by latent periods where the crystal can remain stable for many decades of nanoseconds. The results appear to confirm the process of activated melting in ice, and hence, the likelihood of overheating of the basal face.  Whereas such behavior is far less evident for prism faces of ice, we note that equilibration of the premelting layer thickness close to melting can also take rather long times, and makes it difficult to provide a lower bound of the melting temperature when using a direct coexistence method.

This problem appears to have been the case in most previous studies of ice premelting of the TIP4P/Ice model,  including our own \cite{llombart19,llombart20,llombart20b,sibley21,luengo22b}, where results for premelting film thicknesses were reported for temperatures above the currently accepted melting point just below 270~K \cite{conde17}. Particularly, in our previous work for the TIP4P/Ice we have estimated disjoining pressures and interface potentials for the quasi-liquid layer of ice under the assumption of a melting temperature of 272~K. Acknowledging that some of our results for the average film thickness were obtained actually above melting implies that the disjoining pressures reported previously need to be shifted by about 20~atm, and must therefore adopt negative values for film thicknesses above ca. 9~\AA. This implies that the interface potential of the quasi-liquid layers has necessarily a negative minimum, and confirms the theoretical expectation \cite{luengo22b} that the pristine ice surface exhibits only a limited premelting thickness at the melting point, and is therefore a case of 
{\em incomplete} surface melting. In fact, early reports of the {\em equilibrium} film thickness at temperatures of 270~K can now be interpreted in retrospective as the manifestation of incomplete surface melting for this model \cite{conde08,kling18}. Our calculations suggest that the total surface free energy of the ice/vapor interface at the triple point is barely 0.1~J/m$^2$ above the value corresponding to a situation of complete surface melting, which amounts to the sum of solid/liquid and liquid/vapor surface tensions of ca. 105~J/m$^2$.

In our work, the use of the TIP4P/Ice model developed by Abascal, Sanz, Garc{\'i}a Fern{\'a}ndez and Vega  has been an invaluable tool \cite{abascal05}. Our results on ice premelting in recent years  would have been more accurately interpreted had we relied on an early estimate of its melting temperature coauthored by Abascal \cite{fernandez06}, just to mention one testimony of the care and dedication he exercised in computer simulations.  Some of us (LGM and EGN) had the opportunity to experience this good practice  through a number of enjoyable collaborations \cite{sanz04,sanz04b,macdowell04b,mcbride05,vega05,pi09}. Abascal's account of that period was reviewed in \cite{abascal07c}. LGM recalls  a memorable moment just after finalizing the optimization of  TIP4P parameters that describes this comradeship and friendly atmosphere cherished by Abascal for the research group:  Knocking at his office's door Abascal announced "Habemus par{\'a}metros". 
He would not have guessed, but that day the fate of acclaimed TIP4P and SPC models changed for ever, and so did the careers of most of the members of that research team.

\section{Acknowledgments}

We acknowledge funding from the Spanish Agencia Estatal de Investigaci\'on
under research grant PID2020-115722GB-C21/AEI/10.13039/501100011033, partially sponsored with FEDER funds. We also benefited from generous allocation of computer time at the
Academic Supercomputer Center (CI TASK) in Gdansk.

\section{Appendix}

The interface potential for a film of substance '2', sandwiched between bulk phases of substance '1' and '3' is given by:
\begin{equation}\label{eq:app1}
g_{123}(h) = g_{13}(h) - g_{12}(h) + g_{22}(h) - g_{23}(h) 
\end{equation}
where here, $g_{ij}(h)$ are the surface free energies between two semi-infinite bodies separated by a distance $h$.  For a system interacting via a truncated Lennard-Jones potential with cut-off distance $R_c$, this may be evaluated analytically under  the sharp-kink approximation, yielding \cite{gregorio12}:
\begin{equation}
g_{ij}(h;R_c) = \left \{
\begin{array}{cc}
 U_{ij}(h) - U_{ij}(R_c) - \chi_{ij}(h;R_c) &  h < R_c \\
  & \\
                 0 & h > R_c
\end{array}
\right .
\end{equation}
In the above expression,
\begin{equation}
U_{ij}(h;R_c) = -\frac{\pi}{3}\rho_i\rho_j\sigma_{ij}^4\epsilon_{ij} \left(\frac{\sigma_{ij}}{h}\right)^2
\end{equation}
and 
\begin{equation}\label{eq:app4}
\chi_{ij}(h;Rc) = - \pi \rho_i\rho_j\sigma_{ij}^4\epsilon_{ij} \left(\frac{\sigma_{ij}}{R_c}\right)^4 \left(\frac{R_c}{\sigma_{ij}}-\frac{h}{\sigma_{ij}}\right)^2
\end{equation}
The results of Eq.(\ref{eq:app1}-\ref{eq:app4}) follow from Ref.\cite{gregorio12}. However,
notice that the factor $1/4$ in the third term of the right hand side of Eq.A4 in Ref.\cite{gregorio12} is in error and should be replaced by a factor of $1/2$ instead.

For a premelting film, all phases are made of the same substance, so all $\sigma$ and $\epsilon$ are the same.  The interface potential for the truncated system then follows as:
\begin{equation}
g_{123}(h) = -\frac{\pi}{3}\sigma^4\epsilon(\rho_1-\rho_2)(\rho_3-\rho_2)
\left \{ \left(\frac{\sigma}{h}\right)^2 -  \left(\frac{\sigma}{R_c}\right)^2 - 3 \left(\frac{\sigma}{R_c}\right)^4 \left(\frac{R_c}{\sigma}-\frac{h}{\sigma}\right)^2  \right \}
\end{equation}
for $h<R_c$ and zero otherwise. Notice this result is obtained by neglecting the $r^{-12}$ contribution of the Lennard-Jones potential, which are very small  in the scale of our fit (c.f. $g\propto 1/30 (\sigma/h)^8$ to leading order). 
In practice we found that the fits to the disjoining pressure are hardly affected by the long range contribution, which dominates for $h$ in the scale of decades of nanometers, but is a negligible contribution in the scale of  a few nanometers.

%
%
%


\end{document}